\documentclass[11pt,twoside]{article}

\usepackage{asp2006}
\usepackage{epsf}
\usepackage{lscape}
\usepackage{graphicx}

\begin{document}
\title{Molecular gas and star formation in M81}   
\author{V. Casasola 1, 2, F. Combes 2, G. Galletta 1, and D. Bettoni 3}   
\affil{1 Universit\`a di Padova, 2 Observatoire de Paris-LERMA, 3 INAF-Osservatorio di Padova}    

\begin{abstract} 
We present IRAM 30m observations of the central $\sim$1.6 kpc of 
the spiral M81 galaxy. The molecular gas appears weak and with an unusual excitation physics.  
We discuss a possible link between low CO emission and weak FUV 
surface brightness.
\end{abstract}


\section{Introduction}   
The M81 galaxy, studied at all wavelengths, is the prototype of
CO-poor galaxies. The CO emission appears confined only to the HII regions in   
the spiral arms, and the central region seems devoid of molecular gas.
This lack of CO emission is even more surprising if we consider that, 
in general, interacting galaxies, like M81, has higher CO luminosity than 
non-interacting galaxies.
This non-typical physics of the molecular gas makes 
M81 an optimum candidate to study the X conversion factor between CO and H$_2$
molecules.

\section{Detection and distribution of the molecular gas}
\begin{figure}[ht]
    \centering
		\plottwo{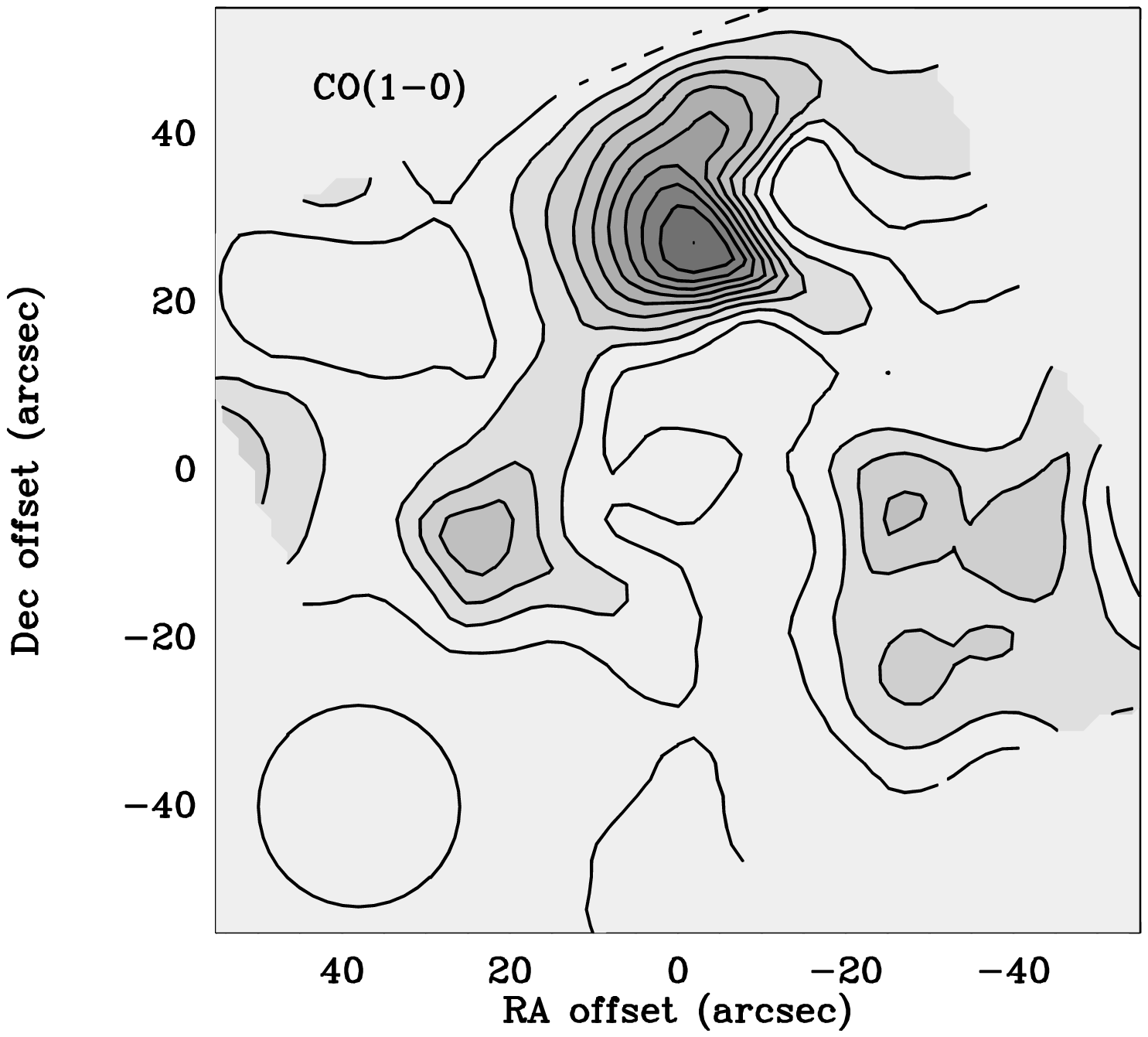}{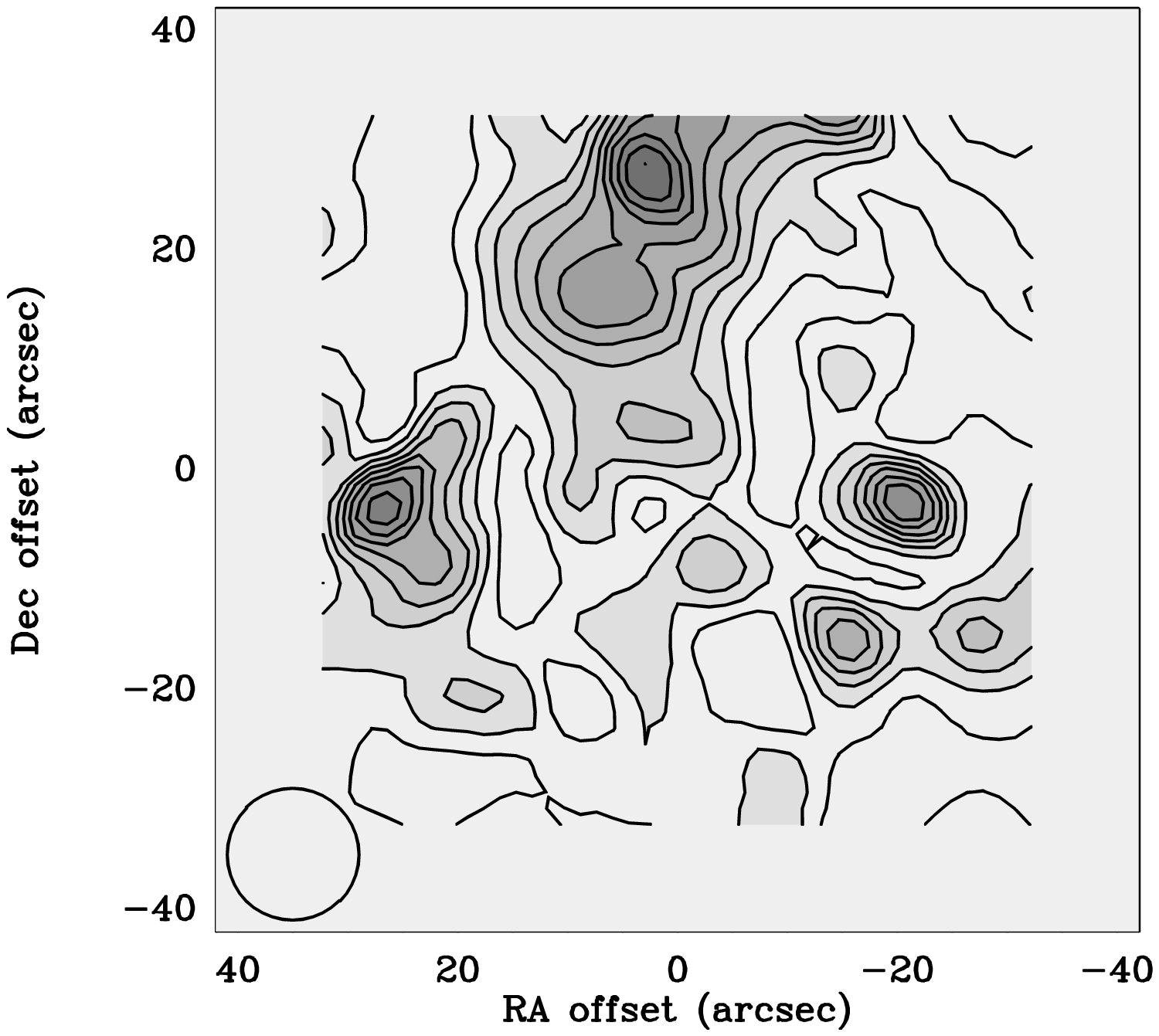}
	  \caption{$\textit{Left}$: Integrated $^{12}$CO(1-0) emission. The contour levels are between 0.701 to
             7.01 K km/s by steps of 0.701 K km/s. The beam of $22^{\prime\prime}$
             is indicated at the bottom left. $\textit{Right}$: Integrated $^{12}$CO(2-1) emission. The contour levels are
            between 0.410 to 4.10 K km/s by steps of 0.410 K km/s. The beam of $12^{\prime\prime}$
            is indicated at the bottom left.}
	  \label{fig2}
   \end{figure}

Observing with the IRAM 30m telescope the inner $\sim$1.6 kpc of M81, we found that 
there are both regions completely devoid of molecular gas
and regions in which the CO is clearly detected and the signal is strong.
In agreement with Brouillet, Baudry, \& Combes (1988) and \citet{sakamoto2}, the central $\sim$300 pc are 
devoid of CO(1-0) emission, and the more important molecular gas emission is a 
clump-structure at the distance of around 460 pc from the nucleus in the North-East direction
(Fig. \ref{fig2}, left panel).

\section{Line ratio and clumping properties}
Studying the line intensities, we found very low temperatures both in CO(1-0)
and CO(2-1) transitions: for the detections $\geq2\sigma$ the average brightness 
temperature in CO(1-0) is of 33.44 mK, while in CO(2-1) is 
of 21.47 mK.  The average $R_{21}$ ratio is of 0.68, a quite low and non-typical 
value for similar galactic nuclei. 
Usually, the inner kpc of galaxies have, on average, $R_{21}\sim$0.9, and it is expected 
even higher for interacting/perturbed galaxies \citep{braine1}. In addition, the line 
ratio typically decreases from galactic nuclei to spiral arms
(Garcia-Burillo, Guelin, \& Cernicharo 1993). For M81 galaxy the $R_{21}$ line ratio spans 
in a wide range of values, without a clear radial trend.

The detailed study of the CO(2-1) emission and distribution, shown in Fig. \ref{fig2} right panel, 
allowed us to resolve individual molecular structures of diameters of $\sim$250 pc and masses 
of  $\sim$10$^{5}$ M$_\odot$. Molecular structures with these dimensions define Giant 
Molecular Associations (GMAs). The $X=N(H_{2})/I_{CO}$ conversion factor computed for all the 
individual resolved GMAs of M81, assumed virialized, takes a  value 
from $\sim$10 to $\sim$18 times larger than the \textit{standard} Galactic value 
\citep[$X=2.3 \times 10^{20}$ mol cm$^{-2}$ (K km s$^{-1}$)$^{-1}$,][]{strong}.

\section{Heating of the gas}
In M81 center, the molecular gas is absent or sub-thermally excited.
This could be due to a flux of cosmic rays and a far-ultraviolet (FUV) surface brightness 
too low to  heat the molecular gas  component of the interstellar medium. 

\section{Conclusions}
M81 appears as a galaxy with an unusual physics of the molecular gas in the central region,
if compared to galaxies with similar distances and morphological type: 
the CO is absent or very weak, the $R_{21}$ line ratio is very low, and
the $X$ conversion factor is extraordinarily large. We interprete there results as due 
to a lack of excitation of the gas, more than the absence of molecular gas.




\end{document}